\documentclass[preprint,12pt,authoryear]{elsarticle}

\usepackage{amssymb}
\usepackage{amsmath}
\usepackage{lineno}

\usepackage{comment}
\usepackage{times}
\usepackage{graphicx}
\usepackage{booktabs}
\usepackage{natbib}
\usepackage{float}
\usepackage{perpage}
\usepackage{pdflscape}
\usepackage{adjustbox}

\MakePerPage{footnote} 

\journal{Planetary and Space Science}

\begin{document}

\begin{frontmatter}

\title{Photothermal Spectroscopy for Planetary Sciences: A Characterization of Planetary Materials in the Mid-IR}

\author[label1,label2]{Christopher Cox} 
\author[label1]{Jakob Haynes} 
\author[label1,label2]{Christopher Duffey} 
\author[label2]{Myles C. Hoskinson} 
\author[label2]{Christopher Bennett} 
\author[label1]{Julie Brisset} 

\address[label1]{Florida Space Institute, University of Central Florida, 12354 Research Pkwy, Orlando FL-32826, USA}
\address[label2]{Department of Physics, University of Central Florida, 4111 Libra Dr, Orlando FL-32816, USA}

\begin{abstract}

Understanding of the formation and evolution of the Solar System requires understanding key and common materials found on and in planetary bodies. Mineral mixing and its implications on planetary body formation is a topic of high interest to the planetary science community. Previous work establishes a case for the use of Optical PhotoThermal InfraRed (O-PTIR) in planetary science and introduces and demonstrates the technique's capability to study planetary materials. In this paper, we performed a measurement campaign on granular materials relevant to planetary science, such as minerals found in lunar and martian soils. These laboratory measurements serve to start a database of O-PTIR measurements. We also present FTIR absorption measurements of the materials we observed in O-PTIR for comparison purposes. We find that the O-PTIR technique suffers from granular orientation effects similar to other IR techniques, but in most cases, is is directly comparable to commonly used absorption spectroscopy techniques. We conclude that O-PTIR would be an excellent tool for the purpose of planetary material identification during in-situ investigations on regolith and bedrock surfaces.

\end{abstract}

\end{frontmatter}

\section{Introduction}
\label{s:intro}

O-PTIR, or Optical-PhotoThermal InfraRed, is a relatively new form of spectroscopy \citep{spadea2021analysis}. It is a pump-probe technique that utilizes an IR laser (pumping laser) and a visible laser (probing laser) \citep{bazin2022using,paulus2022correlative}. The IR laser creates a photothermal effect causing thermal expansion at the surface of a sample and causes a change in the index of refraction of the sample \citep{olson2020simultaneous}. This method is described in detail in \cite{cox2024photothermal}, \cite{bazin2022using}, \cite{krafft2022optical}, and \cite{spadea2021analysis} among many others. \cite{cox2024photothermal} makes a case for the use of O-PTIR technology in the future of planetary sciences. \cite{cox2024photothermal} demonstrates the fascinating capabilities of the O-PTIR technique. Particularly, the case is made for fast measurement speeds while maintaining high spatial and spectral resolution. Additionally, they demonstrate the lack of need for sample preparation and outline the benefits for this technique in planetary science missions. 

The present work will focus on the use of O-PTIR to study minerals relevant to planetary science. Through analysis of returned Apollo samples and x-ray diffraction performed by the Curiosity rover on Mars, we know that materials exist on the surfaces of the Moon and Mars. From these studies, we also know the materials to be granular. For this reason, we decided to study specific minerals found on the Moon and Mars, looking at granular samples in particular.

Detailed mineral composition analysis of regolith grains requires high spatial resolution, ideally at the micron or sub-micron level. To achieve this level of resolution, in-situ measurements or laboratory analysis (via a sample return mission) are usually required. Instrumentation involving Raman (e.g. SHERLOC on Mars Perseverance Rover, \cite{bhartia2021perseverance}, Mass Spectrometry (Cassini's INMS, \cite{waite2004cassini}), and Scanning Electron Microscopes (SEM) (e.g. returned Stardust Samples, \cite{stroud2014stardust}) can produce a spatial resolutions in the micron or even sub-micron range.  However, sometimes these methods are destructive to the sample or may require additional information to identify a material. 

Mineral analysis, especially in situ, yields information regarding the evolution of our Solar System as well as individual planetary bodies \citep{andersen2005carbonaceous,gomes2007minerals}. This includes insight into how materials segregated over time to their current locations \citep{cameron1988origin}. Measurements with sub-micron resolution allow for the ability to distinguish between differing evolutionary histories \citep{ehrenfreund2000organic,tice2022alteration}. 

Given this paper's focus on mineralogical study, the wavenumber range used here is 980 - 1800 $cm^{-1}$. This is primarily due to the key features found there that help unambiguously identify minerals and compounds. \cite{hu2016far} defines the "fingerprint region" to be a region that contains many unique and complex absorptions. It is typically accepted that it lies in the 500 - 1500 $cm^{-1}$ (20.0-6.67 $\mu m$) range. Specifically, there are key features in the 805 - 1300 $cm^{-1}$ (12.4-7.7 $\mu m$) range. An additional limitation is the range of the instrument used.

In this work, we focus on mid-IR wavelengths in order to detect the features that exist within that range for materials relevant to planetary science. Minerals, such as olivine, basalt, anorthosite, etc., are often well characterized in IR wavelengths and specifically in the wavelengths of interest for this work, as evidenced by the existence of multiple openly available databases. These include Infrared and Raman Users Group (IRUG)\footnote{http://www.irug.org/search-spectral-database}, SpectraBase\footnote{https://spectrabase.com/}, and Wiley's Knowitall database\footnote{https://sciencesolutions.wiley.com/knowitall-analytical-edition-software/} among many others. Databases such as these are sometimes open access and they possess entries for materials such as the ones analyzed in this work. Additionally, several minerals are discussed and described in \textit{The Infrared Spectra of Minerals} \citep{farmer1974ism}. The textbook and each above database include data that overlaps with our studied wavelength range, which allows us to include discussions from the textbook into the results discussions. Further, we performed FTIR absorbance (FTIR-A) measurements and utilized Wiley's Knowitall database for direct comparisons of IR absorption measurements to O-PTIR measurements of our samples.

In this paper, we will describe how measurements were taken and analysis of the data produced in Section \ref{s:methods}. We will present our individual measurements by material in Section \ref{s:results}. In Section \ref{s:discuss} we will discuss the ability of O-PTIR to identify and study planetary materials. Finally, in Section \ref{s:conclusion}, we will summarize our work and conclude the paper.

\section{Methodology}
\label{s:methods}

This measurement campaign, using O-PTIR technology, produced mid-IR data for several materials. In this section, we will describe the samples studied, how the measurements were performed, and how the data produced from those measurements were analyzed. The FTIR-A data collection process is also described here.

\subsection{Samples}
\label{s:samples}

Samples in this study are materials relevant to planetary sciences. Below, Table \ref{t:minerals} lists each material that was characterized in this work. The most commonly accepted chemical formula is listed next to the material in the table.

Samples were prepared in the same manner as \cite{cox2024photothermal}.

\begin{table}[H]
  \centering
  \caption{Chemical formulas of each material described and analyzed in this work. Anorthosite's and Basalt's chemical formulas are left out since they are rocks and not a specific mineral.}
    \begin{tabular}{|l|l|}
    \toprule
    \textbf{Mineral} & \textbf{Chemical Formula} \\
    \midrule
    Anorthosite & -- \\
    \midrule
    Basalt & -- \\
    \midrule
    Bronzite & $(Mg,Fe)_2Si_2O_6$ \\
    \midrule
    Fe-Carbonate (Siderite) & $FeCO_3$ \\
    \midrule
    Ferrihydrite & $Fe_{10}O_{14}(OH)_2$ \\
    \midrule
    Gypsum & $CaSO_4\cdot2H_20$ \\
    \midrule
    Hematite & $Fe_2O_3$ \\
    \midrule
    Hydrated Silica & $H_{10}O_3Si$ \\
    \midrule
    Ilmenite & $(Fe,Ti)_2O_3$ \\
    \midrule
    Magnetite & $Fe_3O_4$ \\
    \midrule
    Mg-Carbonate (Magnesite) & $MgCO_3$ \\
    \midrule
    Mg-Sulfate (Epsomite) & $MgSO_4$ \\
    \midrule
    Olivine & $(Mg,Fe)_2SiO_4$ \\
    \midrule
    Smectite & $Al_2H_2O_6Si$ \\
    \bottomrule
    \end{tabular}%
  \label{t:minerals}%
\end{table}%

\subsection{O-PTIR Measurements}
\label{s:measurement}

The calibration and measurement processes are similar to the ones described in \cite{cox2024photothermal}. Here, hyperspectral maps were chosen as the method for data collection. This was decided to help reduce the significance of granular orientation effects. The significance of these effects will be demonstrated in Section \ref{s:results}, where the granular orientation effects for each material will be shown. A hyperspectral map spectra is the average of all hyperspectral points in a given map. Some data presented will be normalized. A normalization factor means every point in the spectrum was multiplied by the factor.

\subsubsection{Hyperspectral Maps}
\label{s:hyperspectral}

In \cite{cox2024photothermal}, it is shown that different materials produced various SNRs. With this knowledge, we elected to measure each material's SNR individually. To do this, we measured each material using single hyperspectral point measurements. One single point measurement is a single spectrum comprised of three scans averaged together by the software of the instrument. Ten such single hyperspectral point measurements were obtained and averaged together to produce a spectrum with error bars. The error bars here are a standard deviation of the ten single hyperspectral point measurements. The spectrum with standard deviation error bars for each material is shown in the right panel of each subsection figure in Section \ref{s:results}. 

\begin{figure}[H]
    \centering
    \includegraphics[width = .85\textwidth]{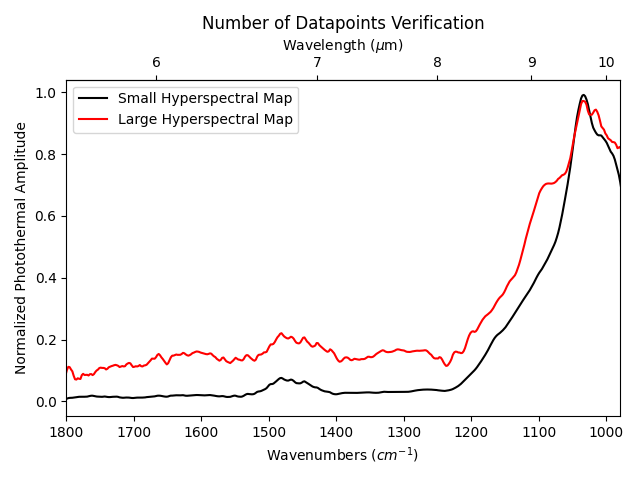}
    \caption{Mid-IR O-PTIR measurements of ilmenite. The solid black line is the measurement discussed in Section \ref{s:ilmenite_results}. The red line is a larger hyperspectral map of the same sample. "Larger" here means a larger measurement area with a similar spacing of measurements. For the smaller hyperspectral map (102 x 76 $\mu m$, 2 $\mu m$ spacing, 2028 individual spectra), instrument settings were: 77\% IR Power, 3.5\% Probe Power, and 20x Detector Gain. The spectrum was normalized by a factor of 1.08 to be more visibly comparable to the larger map. For the larger hyperspectral map (635 x 475 $\mu m$, 5 $\mu m$ spacing, 12288 individual spectra), instrument settings were: 60\% IR Power, 3.5\% Probe Power, and 20x Detector Gain. The spectrum was normalized by a factor of 100 to be more visibly comparable to the smaller hyperspectral map.}
    \label{f:data-ver}
\end{figure}

Figure \ref{f:data-ver} shows the mIRage\textsuperscript{\textregistered} instrument measurements of two separate hyperspectral measurements of the same sample of ilmenite. 

\begin{figure}[H]
    \centering
    \includegraphics[width=\textwidth]{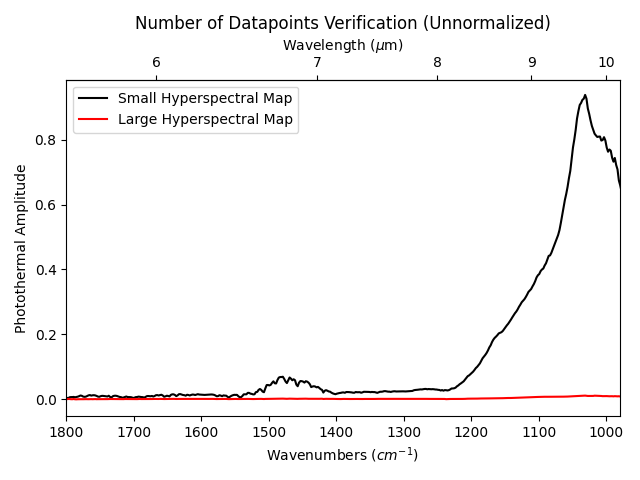}
    \caption{The same Mid-IR O-PTIR measurements of ilmenite shown in Figure \ref{f:data-ver}. This plot removes the normalization. The plot is to serve as a visual aid demonstrating the difference in the measurements.}
    \label{f:data-ver-2}
\end{figure}

The difference between the measurements featured in both Figures~\ref{f:data-ver} and \ref{f:data-ver-2} is the size of the sample probed. The larger map shows a peak at 1016 wavenumbers which is not seen in the spectrum of the smaller hyperspectral map. However, the peaks at 1033, 1448, 1468, and 1481 wavenumbers are all seen in both the spectrum produced by the small map and the spectrum produced by the large map. Figure \ref{f:data-ver} shows the two measurements share very similar spectral peaks and the smaller map spectrum contains most features seen in the larger measurement. The peak of the spectrum of the larger hyperspectral map was much less intense (as seen in Figure \ref{f:data-ver-2}). This is likely due to the increased null space measured between grains. These similarities and the reduction of null space gave us the confidence to use smaller areas to make hyperspectral maps for the purpose of this work.

\subsubsection{Granular Orientation}
\label{s:granular-orientation}

From \cite{cox2024photothermal}, we hypothesize that some of the variation is caused by granular orientation effects. This issue is seen in other more traditional IR methods \citep[e.g.][]{serratosa1958determination,shuai2017quantitative}. In order to study the effect of grain orientation on the spectra we collect, we performed single hyperspectral point measurements (described in Section \ref{s:hyperspectral}) on several different grains in a heterogeneously mixed sample. The found range of spectra from each material is shown in the left panel of each subsection figure in Section \ref{s:results}.

\subsection{FTIR Measurements}
\label{s:ftir-meas}

As stated above, we performed FTIR-A measurements in order to directly compare our O-PTIR data to FTIR data. To collect FTIR measurements, a similar method to \cite{cox2024photothermal} is used. A pellet, prepared as described in Appendix~\ref{a:pellets}, is placed into an iS50R FTIR instrument in a way that the sample is held in front of the beam. The OMNIC software was used to collect the spectra. A pure KBr pellet is placed inside and allowed to purge for a minimum of 12 hours and then scanned as a background reference. We had collection settings set to a gain of 2.0 and an aperture of 87. The measurements were 128 scans with a resolution of 4. Three measurements were taken and their average is reported here.

After background is collected, the pure KBr is removed and replaced with a sample pellet. The mixture consists of the target sample and KBr. Each sample measurement has the background subtracted from it. This subtraction is done by the software. These measurements are used to compare an FTIR measurement with an O-PTIR measurement of the same sample (red curve in the right panel of subsection figures in Section~\ref{s:results}. 

\subsection{Data Analysis}
\label{s:dataanalysis}

To analyze the data, we made hyperspectral maps (described in Section \ref{s:measurement}) and averaged the spectral points to produce a single spectrum. Peaks were identified from this curve using the system's software. Thresholds were set to distinguish peaks from noise in the measurement. Each material's hyperspectral map measurement along with error bars will be presented in their respective sections. The error bars for these plots are the standard error of mean. In a separate plot, the averaged spectrum from the hyperspectral map were normalized to one and compared to absorbance data. The normalization was done by finding the most intense peak and multiplying the entire spectrum by the value required to make the max intensity equal to one.

\section{Results}
\label{s:results}

Using the measurement methods described in Section \ref{s:methods}, we produced mid-IR measurements for several materials relevant to planetary sciences (see a list in Table~\ref{t:run-info}). Spectra are presented as normalized spectra for an easier visual comparison. The "normalization factor" listed in Table \ref{t:run-info} indicates a spectrum was multiplied by that number to normalize the spectrum to one, which summarizes the run information for each measurement discussed in this work.

\newpage
\begin{landscape}

\begin{table}
  \centering
  \caption{Run information for each material analyzed in this work. Each subsection will have this information presented, but it is summarized in this table for ease of reference. "Peak Height" refers to the maximum unnormalized O-PTIR peak measured in the sample, "Normalization Factor" is what each spectrum value was multiplied by to normalize to one for visual comparison purposes, "IR Power" and "Probe Power" indicate the power setting of the respective laser, "HS Map Dimensions" refer to the size in microns of the hyperspectral map made of the sample, "Spacing" refers to the distance in microns between each measurement that composed the map, "Total Spectra" is the total number of measurements in a given map.}
    \begin{adjustbox}{width = 1.5\textwidth}
    \begin{tabular}{|l|r|r|r|r|r|r|r|r|r|r|}
    \toprule
    \textbf{Material} & \multicolumn{1}{l|}{\textbf{Peak Height}} & \multicolumn{1}{l|}{\textbf{Normalization Factor}} & \multicolumn{1}{l|}{\textbf{Max Error}} &  \multicolumn{1}{l|}{\textbf{IR Power (\%)}} & \multicolumn{1}{l|}{\textbf{Probe Power (\%)}} & \multicolumn{1}{l|}{\textbf{Gain (x)}} & \multicolumn{1}{l|}{\textbf{HS Map Dimensions ($\mu m$)}} & \multicolumn{1}{l|}{\textbf{Spacing ($\mu m$)}} & \multicolumn{1}{l|}{\textbf{Total Spectra}} \\
    \midrule
    Anorthosite & 1.113194865 & 0.898315319 & 0.07479 & 46    & 3.5   & 10    & 612 x 468 & 12    & 2080 \\
    \midrule
    Basalt & 0.141632355 & 7.060533591 & 0.02635 & 21    & 3.5   & 20    & 225 x 210 & 5     & 1978 \\
    \midrule
    Bronzite & 13.29053313 & 0.075241526 & 1.72984 & 5     & 3.5   & 10    & 78 x 78 & 2     & 1600 \\
    \midrule
    Fe-Carbonate (Siderite) & 39.08304398 & 0.025586543 & 1.08248 & 10    & 0.43  & 50    & 120 x 84 & 3     & 1189 \\
    \midrule
    Ferrihydrite & 0.015424165 & 64.83333133 & 0.00408 & 60    & 2     & 20    & 588 x 432 & 12    & 1850 \\
    \midrule
    Gypsum & 0.033333541 & 29.9998131 & 0.00956 & 60    & 2     & 10    & 576 x 408 & 12    & 1715 \\
    \midrule
    Hematite & 0.147469541 & 6.781061318 & 0.01356 & 46    & 2     & 20    & 600 x 444 & 12    & 1938 \\
    \midrule
    Hydrated Silica & 15.14671771 & 0.066020904 & 0.77476 & 21    & 3.5   & 10    & 612 x 444 & 12    & 1976 \\
    \midrule
    Ilmenite & 0.938251266 & 1.065812577 & 0.03073 & 77    & 3.5   & 20    & 102 x 76 & 2     & 2028 \\
    \midrule
    Magnetite & 0.805184896 & 1.241950768 & 0.04704 & 46    & 2     & 20    & 588 x 432 & 12    & 1850 \\
    \midrule
    Mg-Carbonate (Magnesite) & 2.506938405 & 0.398892928 & 0.12751 & 21    & 2     & 10    & 612 x 444 & 12    & 1976 \\
    \midrule
    Mg-Sulfate (Epsomite) & 0.105215506 & 9.504302531 & 0.00912 & 21    & 3.5   & 20    & 564 x 432 & 12    & 1776 \\
    \midrule
    Olivine & 3.062116513 & 0.326571506 & 0.17533 & 77    & 2     & 20    & 142 x 70 & 2     & 2592 \\
    \midrule
    Smectite & 10.86572062 & 0.092032552 & 0.57973 & 21    & 3.5   & 5     & 540 x 380 & 10    & 2145 \\
    \bottomrule
    \end{tabular}%
    \end{adjustbox}
  \label{t:run-info}%
\end{table}%

\end{landscape}
\newpage

In this section, we discuss the results of each material listed (in alphabetical order). The findings are described and displayed in each subsection, but a summary is provided in Table \ref{t:mineral_peaks}. For each material we show a figure containing a left and right plot. The left plot presents the granular orientation effects seen in the measurements of individual material grains. Each line is the average of a set of ten O-PTIR single point hyperspectral measurements (described in Section \ref{s:hyperspectral}). The spectra are recorded on different grains, which naturally have different orientations in the sample, leading to the different features in their spectra. The gray error bars represent the standard deviation of those ten measurements. In these plots, the coloration does not matter other than to distinguish between the spectra. Given that some of the spectra showed features with very low intensities, we chose to show them in a log-lin plot to make these visible compared to more dominating features.

The right plot of each materials' figure presents a comparison between an O-PTIR measurement (black line), an FTIR-A measurement (red line), and a comparable Wiley's Knowitall database entry (cyan line). In this plot, the O-PTIR measurement is an average over a hyperspectral map. The O-PTIR data utilizes the left y-axis and the comparisons utilize the right y-axis. 

\begin{table}[H]
  \centering
  \caption{Peak wavenumbers and wavelengths for each  material examined in this work. Dashes in a cell indicate the lack of an O-PTIR peak measured in this wavenumber range for that material.}
    \begin{tabular}{|l|l|l|}
    \toprule
    \textbf{Mineral} & \textbf{Peak(s) (Wavenumber, $cm^{-1}$)} & \textbf{Peak(s) (Wavelength, $\mu m$)} \\
    \midrule
    Anorthosite & 986, 1027, 1150 & 10.1, 9.7, 8.7 \\
    \midrule
    Basalt & 1004, 1447, 1465, 1483 & 10.0, 6.9, 6.8, 6.7 \\
    \midrule
    Bronzite & 1023, 1465 & 9.8, 6.8 \\
    \midrule
    Fe-Carbonate (Siderite) & 1039, 1444, 1465 & 9.6, 6.9, 6.8 \\
    \midrule
    Ferrihydrite & --    & -- \\
    \midrule
    Gypsum & 990, 1007, 1108, 1622 & 10.1, 9.9, 9.0, 6.2 \\
    \midrule
    Hematite & 1013, 1020, 1087 & 9.9, 9.8, 9.2 \\
    \midrule
    Hydrated Silica & 1082, 1632 & 9.2, 6.1 \\
    \midrule
    Ilmenite & 1033, 1448, 1468, 1481 & 9.7, 6.9, 6.8, 6.7 \\
    \midrule
    Magnetite & --    & -- \\
    \midrule
    Mg-Carbonate (Magnesite) & 1021, 1468, 1482, 1514 & 9.8, 6.8, 6.7, 6.6 \\
    \midrule
    Mg-Sulfate (Epsomite) & 1077, 1101, 1665 & 9.3, 9.1, 6.0 \\
    \midrule
    Olivine & 1006, 1125 & 9.9, 8.9 \\
    \midrule
    Smectite & 1028, 1447, 1466 & 9.7, 6.9, 6.8 \\
    \bottomrule
    \end{tabular}%
  \label{t:mineral_peaks}%
\end{table}%

\subsection{Anorthosite}
\label{s:anorthosite}

\begin{figure}[H]
    \centering
    \includegraphics[width=\textwidth]{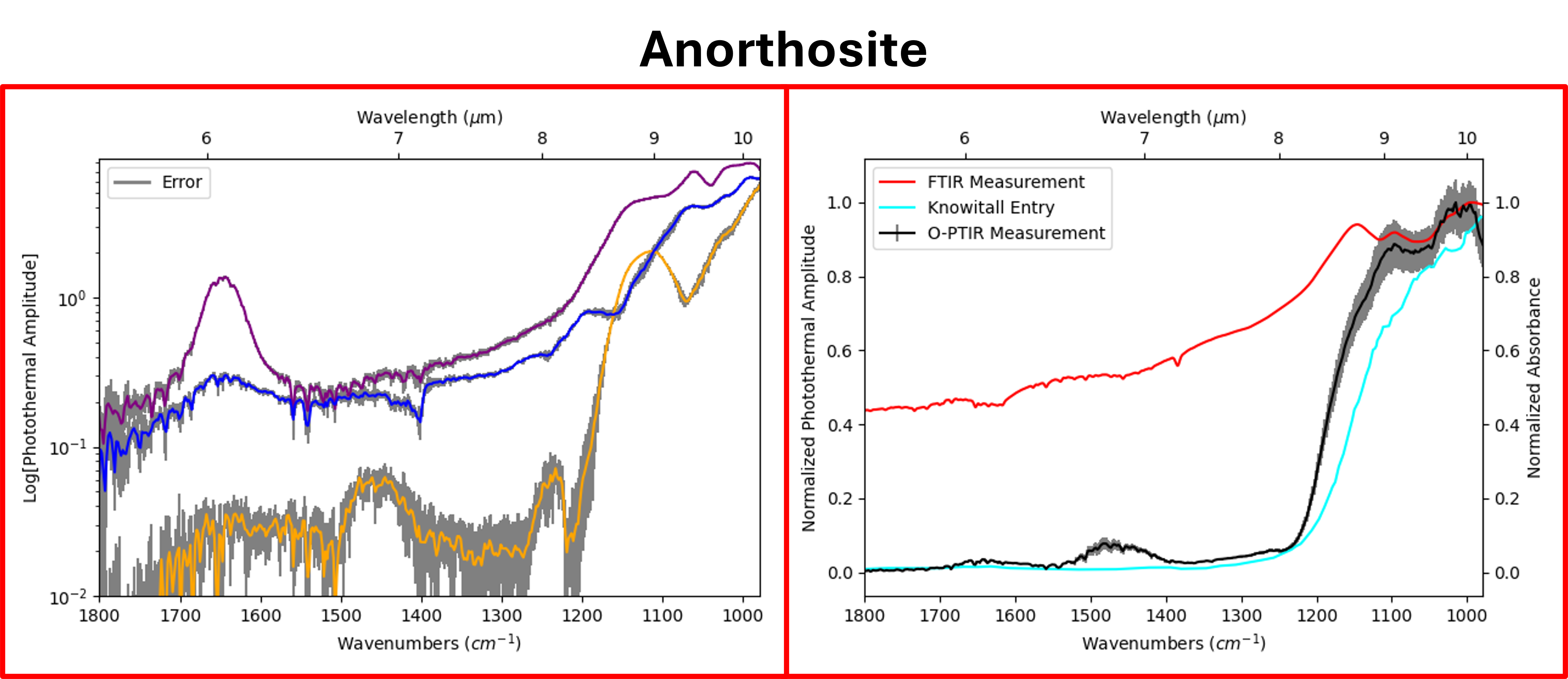}
    \caption{Mid-IR O-PTIR measurements of anorthosite (left) and the average spectrum from an O-PTIR hyperspectral map of anorthosite (right).}
    \label{f:anorthosite-figure}
\end{figure}

Figure \ref{f:anorthosite-figure} (left) shows O-PTIR measurements of anorthosite at chosen locations on the sample to show variations we hypothesize are due to grain orientation.

Figure \ref{f:anorthosite-figure} (right) shows an O-PTIR measurement of anorthosite with standard error of mean error bars. Anorthosite shows O-PTIR peaks at 998, 1019, and 1095 wavenumbers. The double peak feature in the lower wavenumber range was seen in two of the three spectra featured in Figure \ref{f:anorthosite-figure} (left). The O-PTIR measurements compares quite well to both the FTIR measurement of the same sample (Figure \ref{f:anorthosite-figure}, right, red line) as it shares similar peak locations and a similar spectral shape. Additionally, the O-PTIR spectrum matches the anorthite spectrum (Figure \ref{f:anorthosite-figure}, right, cyan line) quite well given the same double feature and matching peak locations. We made the comparison between anorthosite and anorthite due to the similar composition despite anorthosite being a plagioclase \textit{rock} and anorthite being a plagioclase \textit{mineral}.

This also is consistent with \cite{farmer1974ism} which lists peaks associated with anorthite in the IR. In the examined wavenumber range, anorthite peaks at 1020, 1085, and 1160. The peaks at 1020 and 1085 correspond closely with two of the measured O-PTIR peaks.

\subsection{Basalt (Glass-Rich)}

\begin{figure}[H]
    \centering
    \includegraphics[width=\textwidth]{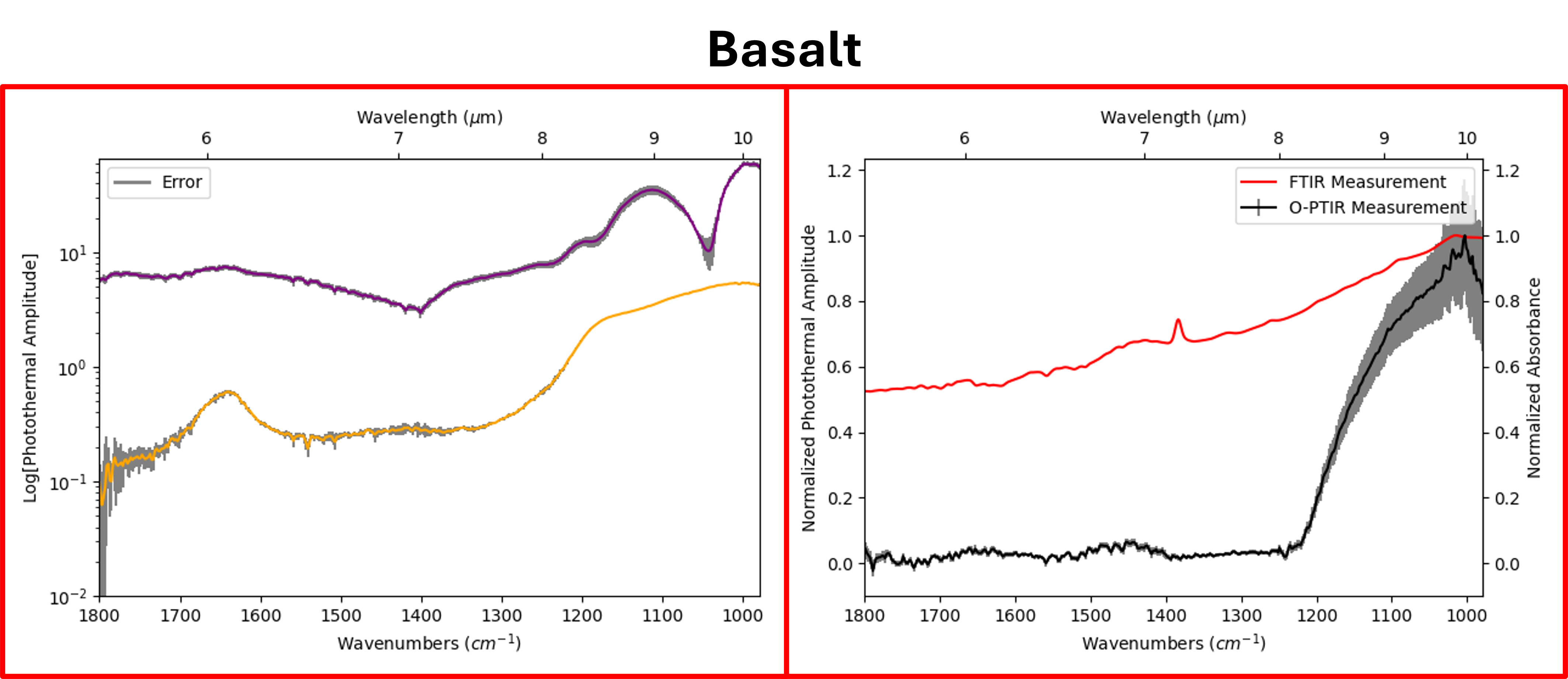}
    \caption{Mid-IR O-PTIR measurements of basalt (left) and the average spectrum from an O-PTIR hyperspectral map of basalt (right).}
    \label{f:basalt-figure}
\end{figure}

Figure \ref{f:basalt-figure} (left) shows O-PTIR measurements of basalt at chosen locations on the sample to show variations we hypothesize are due to grain orientation. 

Figure \ref{f:basalt-figure} (right) shows an O-PTIR measurement of basalt with standard error of mean error bars. Basalt shows O-PTIR peaks at 1004, 1447, 1465, and 1483 wavenumbers. The O-PTIR spectrum shared a major peak with the FTIR measurement of the same sample. Though the spectral shape is not a match, there is significant decrease in signal for both spectra following the max peak. There were no samples of close enough composition for basalt in the Wiley Knowitall database.

\subsection{Bronzite}
\label{s:bronzite-results}

\begin{figure}[H]
    \centering
    \includegraphics[width=\textwidth]{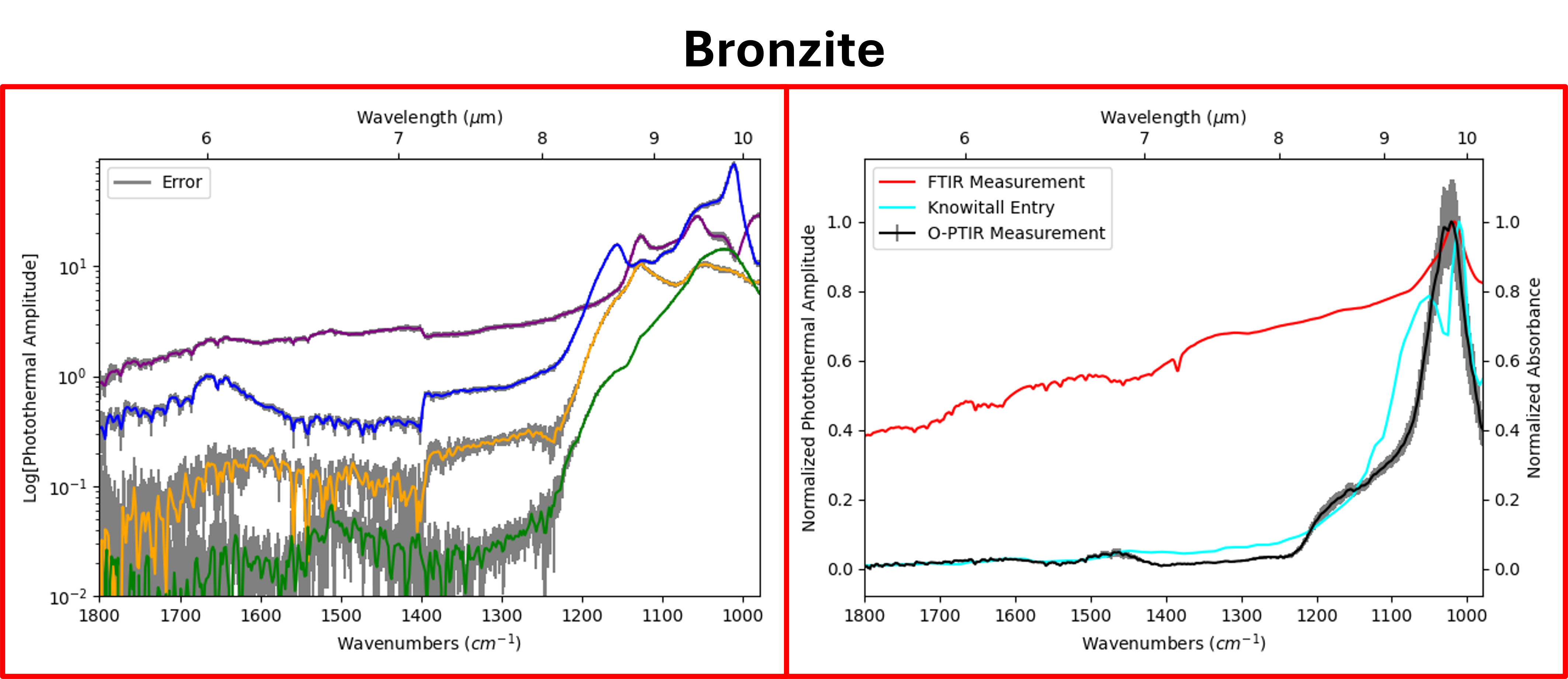}
    \caption{Mid-IR O-PTIR measurements of bronzite (left) and the average spectrum from an O-PTIR hyperspectral map of bronzite (right).}
    \label{f:bronzite-figure}
\end{figure}

Figure \ref{f:bronzite-figure} (left) shows O-PTIR measurements of bronzite at chosen locations on the sample to show variations we hypothesize are due to grain orientation.

Figure \ref{f:bronzite-figure} (right) shows an O-PTIR measurement of bronzite with standard error of mean error bars. Bronzite shows O-PTIR peaks at 1023 and 1465 wavenumbers. The averaged hyperspectral map spectrum appears nearly identical to one of the granular orientation effect images (Figure \ref{f:bronzite-figure}, left, cyan line). This spectrum had the most intense peak among the whole set of granular orientation effect spectra for bronzite. The O-PTIR spectrum shared a peak with both the FTIR measurement of the same sample (red line) and a bronzite entry in the Wiley Knowitall database. Even though the spectral shape between the O-PTIR measurement and the FTIR measure of the same sample do not match, there is a consistent decrease in both measurements following the major peak. Additionally, the O-PTIR measurement closely resembles the database entry for bronzite despite the spectral feature around the peak not being identical.

\subsection{Fe-Carbonate (Siderite)}

\begin{figure}[H]
    \centering
    \includegraphics[width=\textwidth]{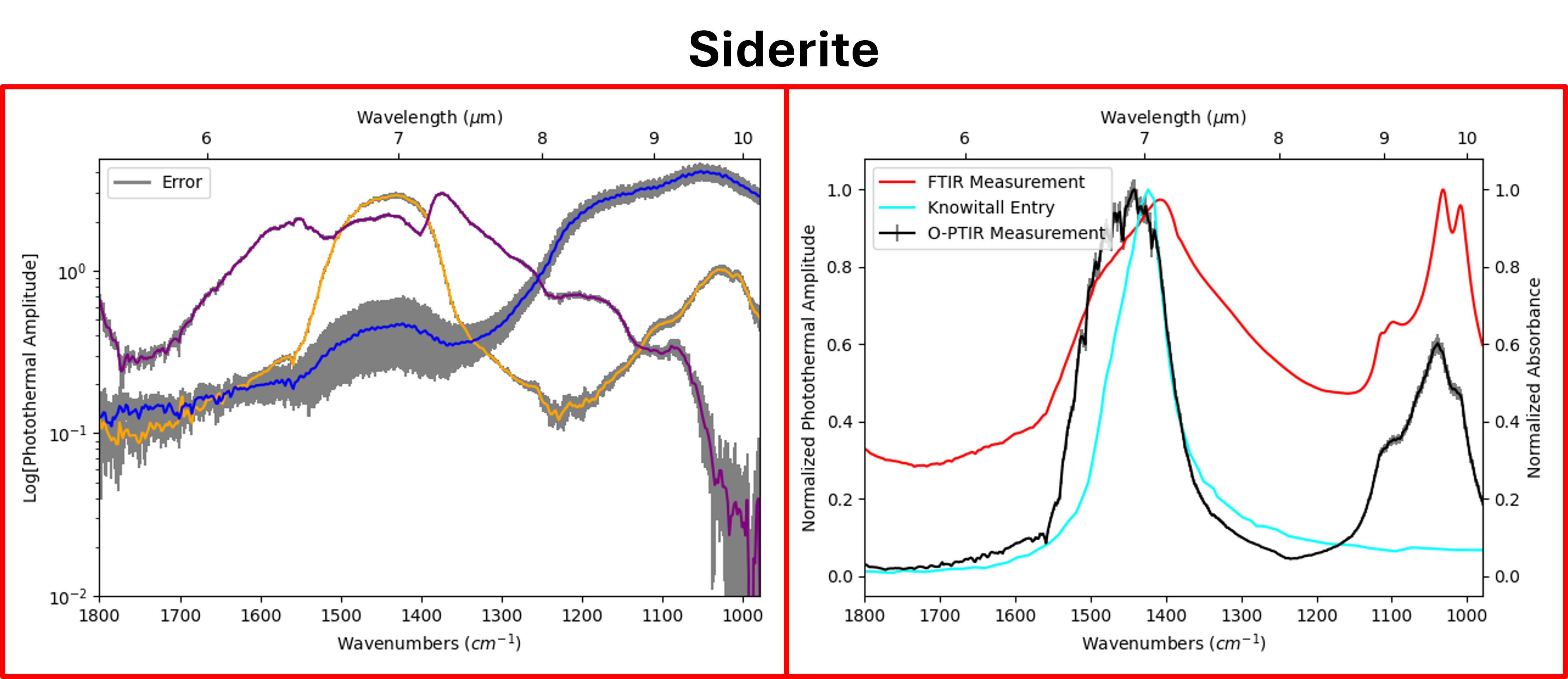}
    \caption{Mid-IR O-PTIR measurements of siderite (left) and the average spectrum from an O-PTIR hyperspectral map of siderite (right).}
    \label{f:siderite-figure}
\end{figure}

Figure \ref{f:siderite-figure} (left) shows O-PTIR measurements of siderite at chosen locations on the sample to show variations we hypothesize are due to grain orientation. 

Figure \ref{f:siderite-figure} (right) shows an O-PTIR measurement of siderite with standard error of mean error bars. Siderite shows O-PTIR peaks at 1039, 1444, and 1465 wavenumbers. The O-PTIR measurement shares peaks with both the FTIR measurement of the same sample and the Knowitall database entry for siderite. The O-PTIR feature between 1000 and 1150 $cm^{-1}$ is not identical to but does match the FTIR feature on the same sample. This feature does not exist in the Knowitall database entry. The O-PTIR feature between 1350 and 1550 $cm^{-1}$ has similar peaks in both the FTIR measurement of the same sample and the Knowitall database entry. For the the FTIR measurement of the same sample, the peak is shifted to a lower wavenumber, but is close. The Knowitall database entry peaks at the same wavenumber but the feature is not nearly as wide. The FTIR measurement of the same sample has a very similar spectral shape to the O-PTIR measurement despite the shifted peak at the higher wavenumber.

According to \cite{farmer1974ism}, siderite has absorption features at $1071$, $1415$, and $1422$. The siderite O-PTIR measured at 1039 $cm^{-1}$ could be attributed to the $1071$ feature. It is 42 wavenumbers displaced, but the width of that feature is approximately 200 wavenumbers, so it could cover that band. Additionally, the peaks at 1444 and 1465 $cm^{-1}$ are part of a feature with a width of almost 300 wavenumbers. This puts this peak well within the range of values listed for $v_3$ for siderite.

\cite{farmer1974ism} also lists peaks associated with siderite in the IR. In the examined wavenumber range, siderite peaks at 1412. This peak is only 32 wavenumbers displaced from one of the measured peaks in the O-PTIR. Additionally, the feature around the peak at 1039 is similar in FTIR, but with differing intensities. We hypothesize that this is potentially a polarization issue, but that kind of consideration is beyond the scope of this work.

\subsection{Ferrihydrite}

\begin{figure}[H]
    \centering
    \includegraphics[width=\textwidth]{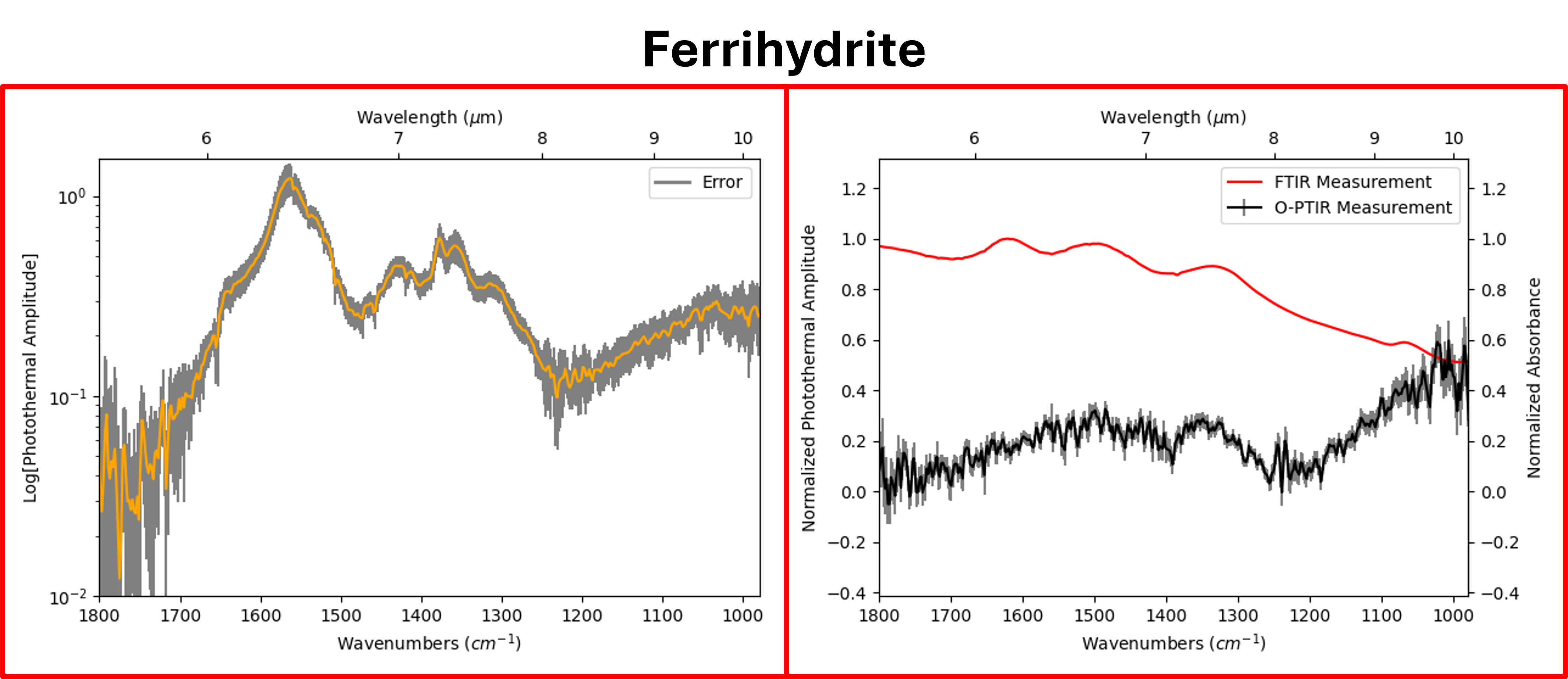}
    \caption{Mid-IR O-PTIR measurements of ferrihydrite (left) and the average spectrum from an O-PTIR hyperspectral map of ferrihydrite (right).}
    \label{f:ferrihydrite-figure}
\end{figure}

A single response was detected for ferrihydrite (shown in Figure \ref{f:ferrihydrite-figure}, left). \cite{cox2024case} was unable to detect any signal from this material. The detected signal was weak and the deviation was high at multiple points in the spectrum. This makes it difficult to evaluate how granular orientation affects ferrihydrite. 

Figure \ref{f:ferrihydrite-figure} (right) shows an O-PTIR measurement of ferrihydrite with standard error of mean error bars. Ferrihydrite does not show significant O-PTIR spectral features nor peaks in the chosen wavenumber range. Though a spectral response was detected on a single grain, the signal was not strong enough to avoid being dominated by the null measurements contained in the sample. The FTIR measurement of the same sample was also devoid of features in this wavenumber range. There were no found entries for ferrihydrite in the Wiley Knowitall database.

\subsection{Gypsum}

\begin{figure}[H]
    \centering
    \includegraphics[width=\textwidth]{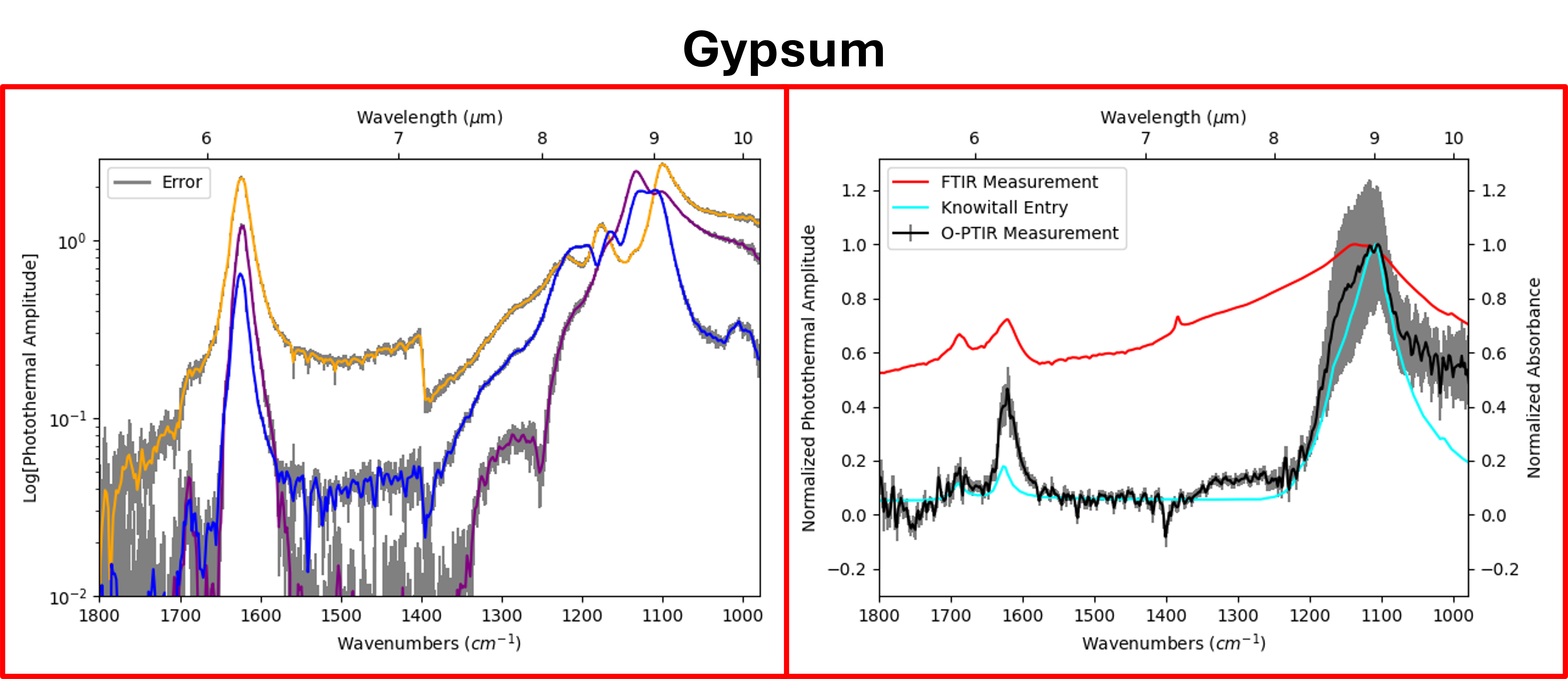}
    \caption{Mid-IR O-PTIR measurements of gypsum (left) and the average spectrum from an O-PTIR hyperspectral map of gypsum (right).}
    \label{f:gypsum-figure}
\end{figure}

Figure \ref{f:gypsum-figure} (left) shows O-PTIR measurements of gypsum at chosen locations on the sample to show variations we hypothesize are due to grain orientation. The sudden increase at 1400 $cm^{-1}$ in Figure \ref{f:gypsum-figure} (left, black line) is caused by a switch between the QCL chips.

The O-PTIR measurement shares peaks and features with both the FTIR measurement of the same sample (Figure \ref{f:gypsum-figure}, right, red line) and the Knowitall database comparison (Figure \ref{f:gypsum-figure}, right, cyan line). There is a double peak in the FTIR measurement of the same sample which may also exist in the O-PTIR measurement, though the second peak in the O-PTIR measurement is noisier. The Knowitall database comparison is nearly identical though the peak at the higher wavenumber is slightly shifted. Additionally, the feature near 1100 wavenumbers is not as wide in the Knowitall entry as it is in the O-PTIR measurement.

According to \cite{farmer1974ism}, anhydrite has absorption features at $1013,$ $1095,$ $1126,$ and $1149$ and gypsum has absorption features at $1000,$ $1006,$ $1117,$ $1118,$ $1131,$ $1138,$ $1142,$ and $1144$. The gypsum O-PTIR measured at 1007 $cm^{-1}$ could be attributed to the feature at $1006$ as it is one wavenumber displaced. The peak at 1108 $cm^{-1}$ is part of a feature with a width of approximately 100 wavenumbers. This puts this peak well within the range of values listed for features for gypsum.

The feature around 1629 is seen in both O-PTIR and FTIR-A, but with differing intensities. This is potentially a polarization issue, but that is beyond the scope of this work.

\subsection{Hematite}

\begin{figure}[H]
    \centering
    \includegraphics[width=\textwidth]{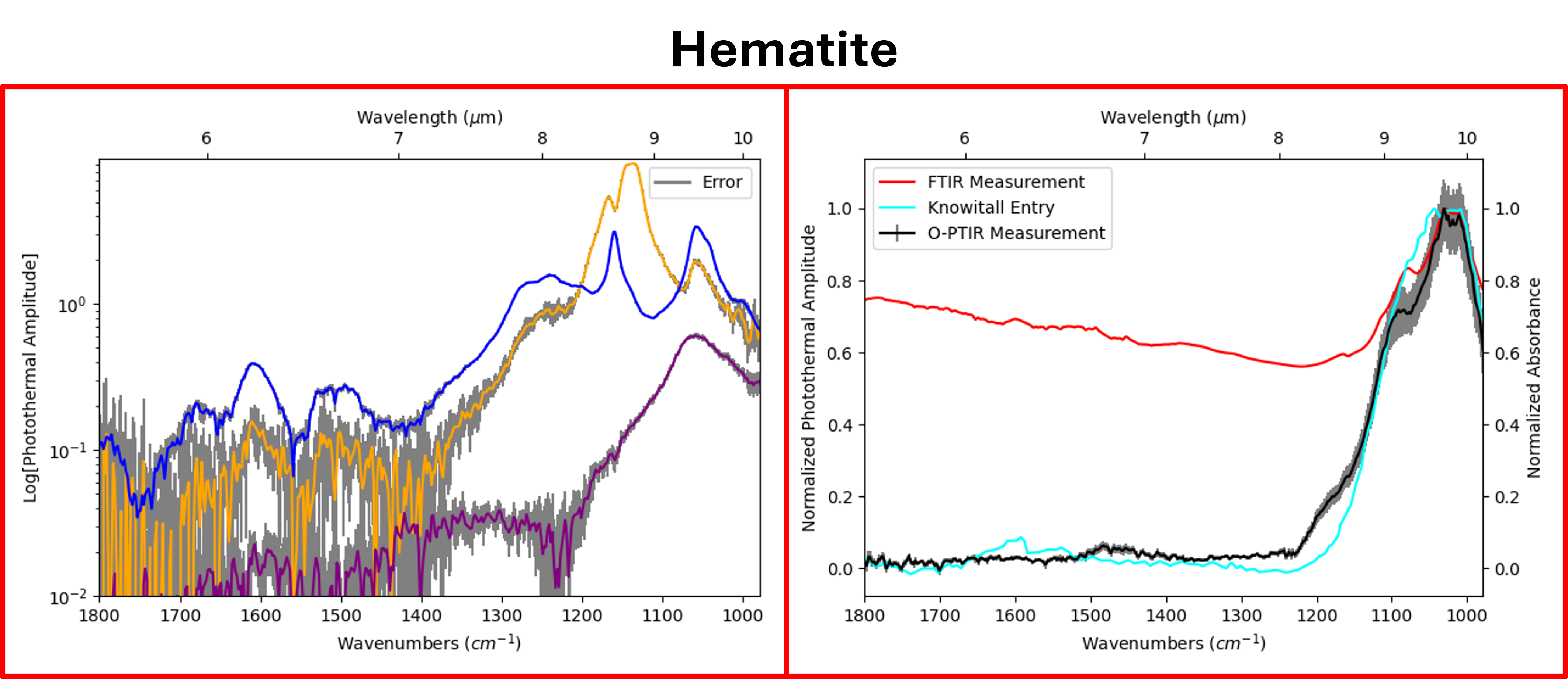}
    \caption{Mid-IR O-PTIR measurements of hematite (left) and the average spectrum from an O-PTIR hyperspectral map of hematite (right).}
    \label{f:hematite-figure}
\end{figure}

Figure \ref{f:hematite-figure} (left) shows O-PTIR measurements of hematite at chosen locations on the sample to show variations we hypothesize are due to grain orientation.

Figure \ref{f:hematite-figure} (right) shows an O-PTIR measurement of hematite with standard error of mean error bars. Hematite shows O-PTIR peaks at 1013, 1020, and 1087 wavenumbers. The O-PTIR measurement shares peaks with both the FTIR measurement of the same sample and the Knowitall database entry for hematite. The Knowitall database entry (\ref{f:hematite-figure}, right, cyan line) does not contain the dip on the left side of the feature that both the FTIR measurement of the same sample and the O-PTIR measurement possess. However, the feature in the Knowitall database entry does have the same width as the other two entries in the figure. The O-PTIR measurement and the FTIR measurement of the same sample are nearly identical in spectral shape though the intensities do vary slightly. All three measurements are devoid of spectral features past 1100 $cm^{-1}$.

\subsection{Hydrated Silica}

\begin{figure}[H]
    \centering
    \includegraphics[width=\textwidth]{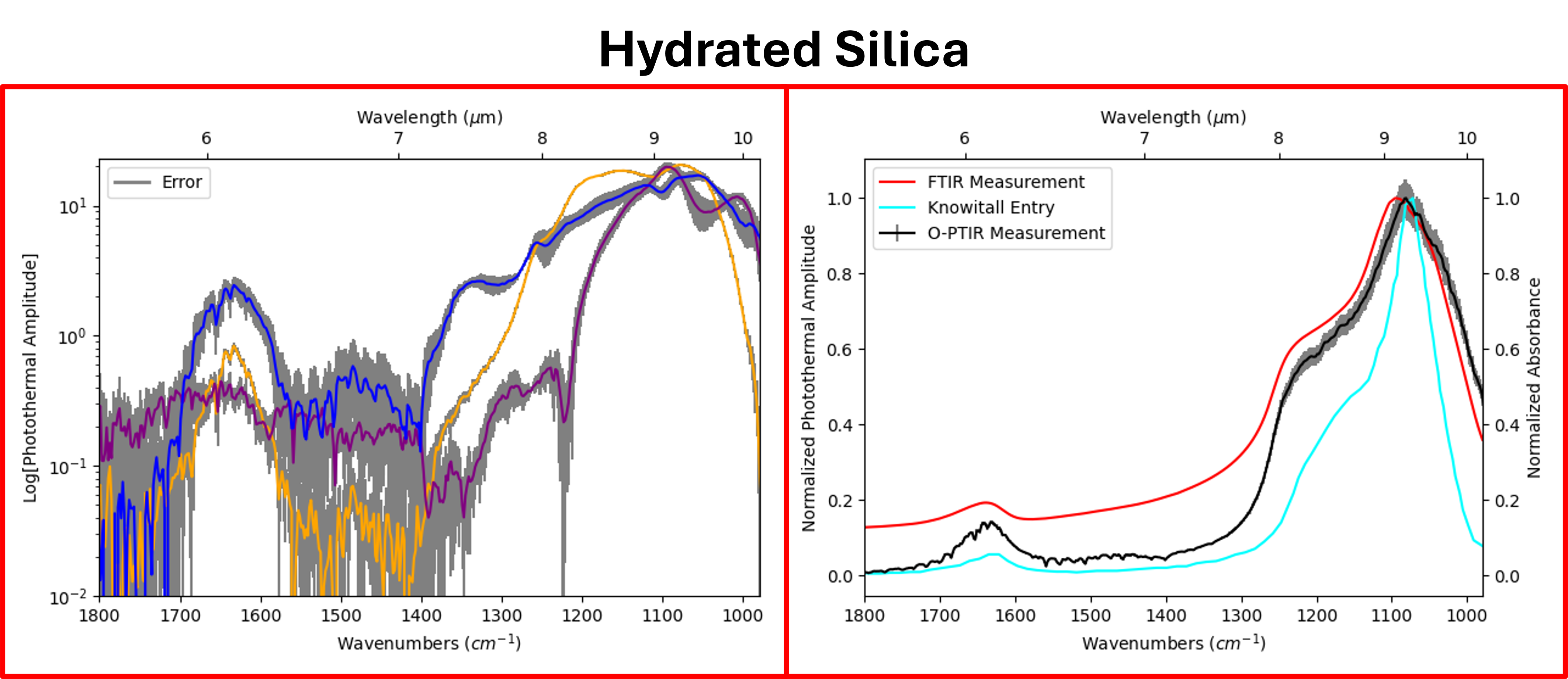}
    \caption{Mid-IR O-PTIR measurements of hydrated silica (left) and the average spectrum from an O-PTIR hyperspectral map of hydrated silica (right).}
    \label{f:hydratedsilica-figure}
\end{figure}

Figure \ref{f:hydratedsilica-figure} (left) shows O-PTIR measurements of hydrated silica at chosen locations on the sample to show variations we hypothesize are due to grain orientation. 

Figure \ref{f:hydratedsilica-figure} (right) shows an O-PTIR measurement of hydrated silica with standard error of mean error bars. Hydrated Silica shows O-PTIR peaks at 1082 and 1632 wavenumbers. The spectral shape of the O-PTIR measurement is extremely similar to both the FTIR measurement of the same sample and the Knowitall entry for hydrated silica. It is nearly identical to the FTIR measurement. In both comparisons the peaks and features all line up.

\subsection{Ilmenite}
\label{s:ilmenite_results}

\begin{figure}[H]
    \centering
    \includegraphics[width=\textwidth]{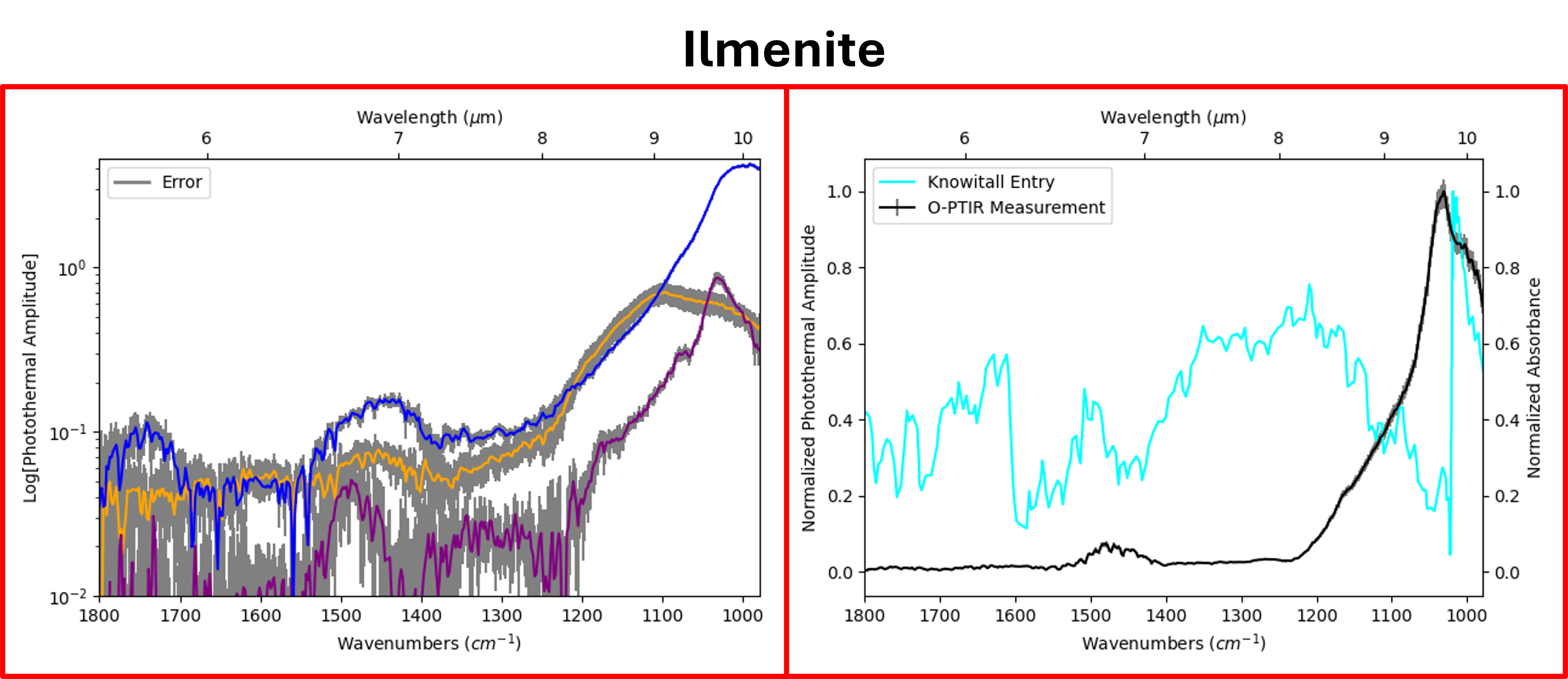}
    \caption{Mid-IR O-PTIR measurements of ilmenite (left) and the average spectrum from an O-PTIR hyperspectral map of ilmenite (right).}
    \label{f:ilmenite-figure}
\end{figure}

Figure \ref{f:ilmenite-figure} (left) shows O-PTIR measurements of ilmenite at chosen locations on the sample to show variations we hypothesize are due to grain orientation. 

Figure \ref{f:ilmenite-figure} (right) shows an O-PTIR measurement of ilmenite with standard error of mean error bars. Ilmenite shows O-PTIR peaks at 1033, 1448, 1468, and 1481 wavenumbers. The averaged hyperspectral map spectrum appears to be most similar to the spectrum represented by the red line in Figure \ref{f:ilmenite-figure} (left). The comparable spectrum has a low intensity especially compared to the spectrum represented by the cyan line in the granular orientation effect figure. The O-PTIR spectrum is not extremely comparable to the available Knowitall database spectrum. However, the two spectra share a peak in the lower wavenumber range. No FTIR measurement of the same sample is presented here because our FTIR-A measurement for Ilmenite was null. This is not uncommon in FTIR-A and likely means the sample was optically thick to the pellet size we produced.

\subsection{Magnetite}

\begin{figure}[H]
    \centering
    \includegraphics[width=\textwidth]{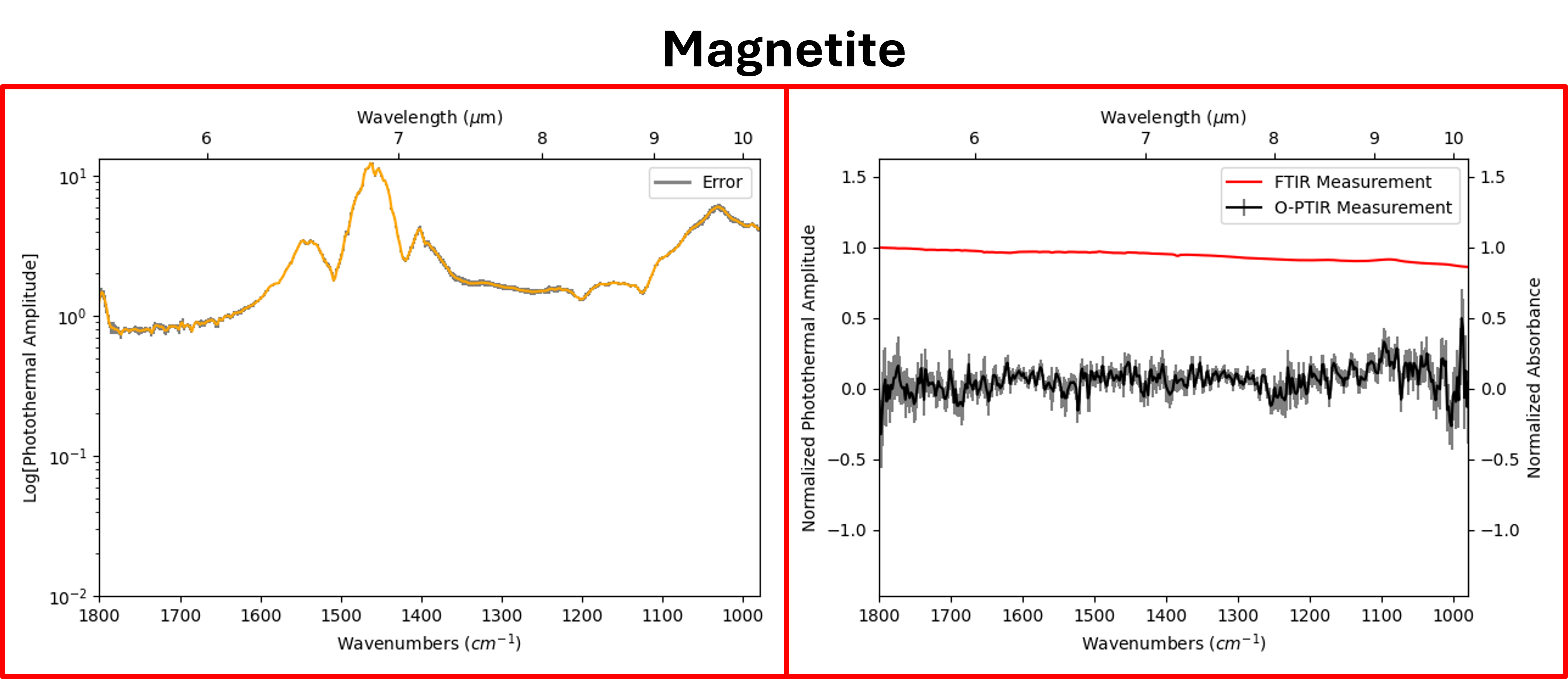}
    \caption{Mid-IR O-PTIR measurements of magnetite (left) and the average spectrum from an O-PTIR hyperspectral map of magnetite (right).}
    \label{f:magnetite-figure}
\end{figure}

Figure \ref{f:magnetite-figure} (left) shows O-PTIR measurements of magnetite at chosen locations on the sample to show variations we hypothesize are due to grain orientation. \cite{cox2024case} was unable to detect any signal from this material. The detected signal was weak and the deviation was high at multiple points in the spectrum. This makes it difficult to evaluate how granular orientation affects magnetite.

Figure \ref{f:magnetite-figure} (right) shows an O-PTIR measurement of magnetite with standard error of mean error bars. Magnetite does not show significant O-PTIR spectral features nor peaks in the chosen wavenumber range. Though a spectral response was detected on a single grain, the signal was not strong enough to avoid being dominated by the null measurements contained in the sample. The FTIR measurement of the same sample was also devoid of features in this wavenumber range. There were no found entries for magnetite in the Wiley Knowitall database. \cite{cox2024case} measured this same sample. In that work, using a similar method, there were also no peaks nor features detected. Figure \ref{f:magnetite-figure} (right) is similar to those findings. 

\subsection{Mg-Carbonate (Magnesite)}

\begin{figure}[H]
    \centering
    \includegraphics[width=\textwidth]{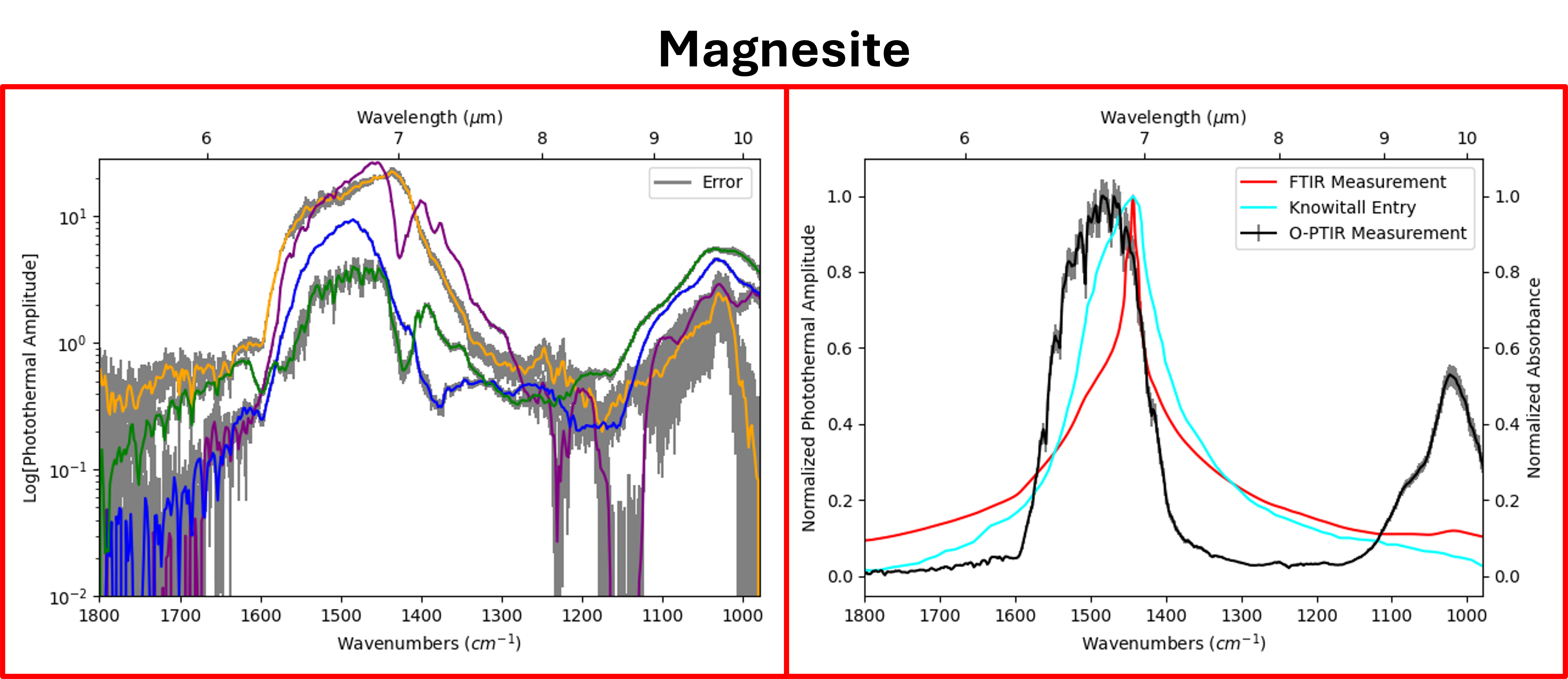}
    \caption{Mid-IR O-PTIR measurements of magnesite (left) and the average spectrum from an O-PTIR hyperspectral map of magnesite (right).}
    \label{f:magnesite-figure}
\end{figure}

Figure \ref{f:magnesite-figure} (left) shows O-PTIR measurements of magnesite at chosen locations on the sample to show variations we hypothesize are due to grain orientation. 

Figure \ref{f:magnesite-figure} (right) shows an O-PTIR measurement of magnesite with standard error of mean error bars. Magnesite shows O-PITR peaks at 1021, 1468, 1482, and 1514 wavenumbers. The averaged hyperspectral map spectrum appears to be most similar to the spectra represented by the black and cyan lines in Figure \ref{f:magnesite-figure} (left). The O-PTIR measurement is comparable with both the FTIR measurement of the same sample as well as the entry in the Knowitall database. All three measurements peak in the higher wave number range at very similar positions. However, the O-PTIR measurement has a wider feature at that peak than both the FTIR measurement of the same sample and the Knowitall database entry. The width of the spectral feature for the O-PTIR measurement is more comparable to the Knowitall database entry.

According to \cite{farmer1974ism}, magnesite has absorption features at $1436$, $1450$, and $1599$. The magnesite O-PTIR measured at 1468 $cm^{-1}$ could be attributed to the feature at $1450$. It is less than 20 wavenumbers displaced, and the width of that feature is approximately 200 wavenumbers, so it could cover that band. The features located at 1436 and 1450 could help account for the concentration of absorbance peaks and the width of the O-PTIR feature at the wavenumber range. The O-PTIR feature around 1021 does not seem to have a corresponding absorbance peak.

\subsection{Mg-Sulfate (Epsomite)}

\begin{figure}[H]
    \centering
    \includegraphics[width=\textwidth]{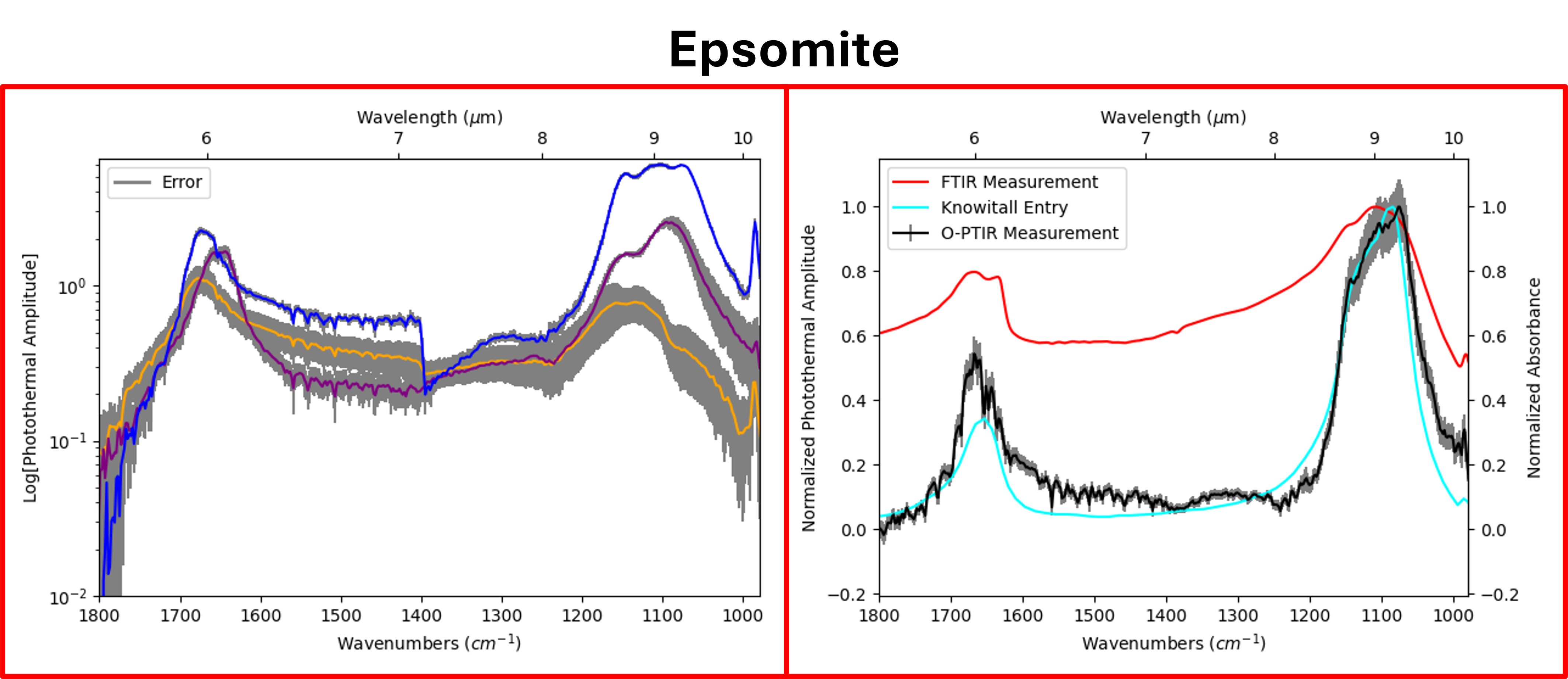}
    \caption{Mid-IR O-PTIR measurements of epsomite (left) and the average spectrum from an O-PTIR hyperspectral map of epsomite (right).}
    \label{f:epsomite-figure}
\end{figure}

Figure \ref{f:epsomite-figure} (left) shows O-PTIR measurements of epsomite at chosen locations on the sample to show variations we hypothesize are due to grain orientation. Figure \ref{f:epsomite-figure} (right) shows an O-PTIR measurement of epsomite with standard error of mean error bars. Epsomite shows O-PTIR peaks at 1077, 1101, and 1665 wavenumbers. The O-PTIR measurement shares similarities with both the FTIR measurement of the same sample and the Knowitall database entry, with an almost perfect match to the Knowitall data and shared peak locations, if not shapes, with the FTIR measurement.

According to \cite{farmer1974ism}, epsomite has absorption features at $1020$, $1085$, $1110$, $1155$, $1175$, and $1235$. The epsomite O-PTIR measured at 1077 $cm^{-1}$ could be attributed to the $1085$ feature as it is less than 10 wavenumbers displaced. Additionally, the peak at 1101 $cm^{-1}$ could be attributed to the $1110$ feature as it is also less than 10 wavenumbers displaced. The feature those peaks are a part of is approximately 200 wavenumbers wide which could cover additional features.

\subsection{Olivine}

\begin{figure}[H]
    \centering
    \includegraphics[width=\textwidth]{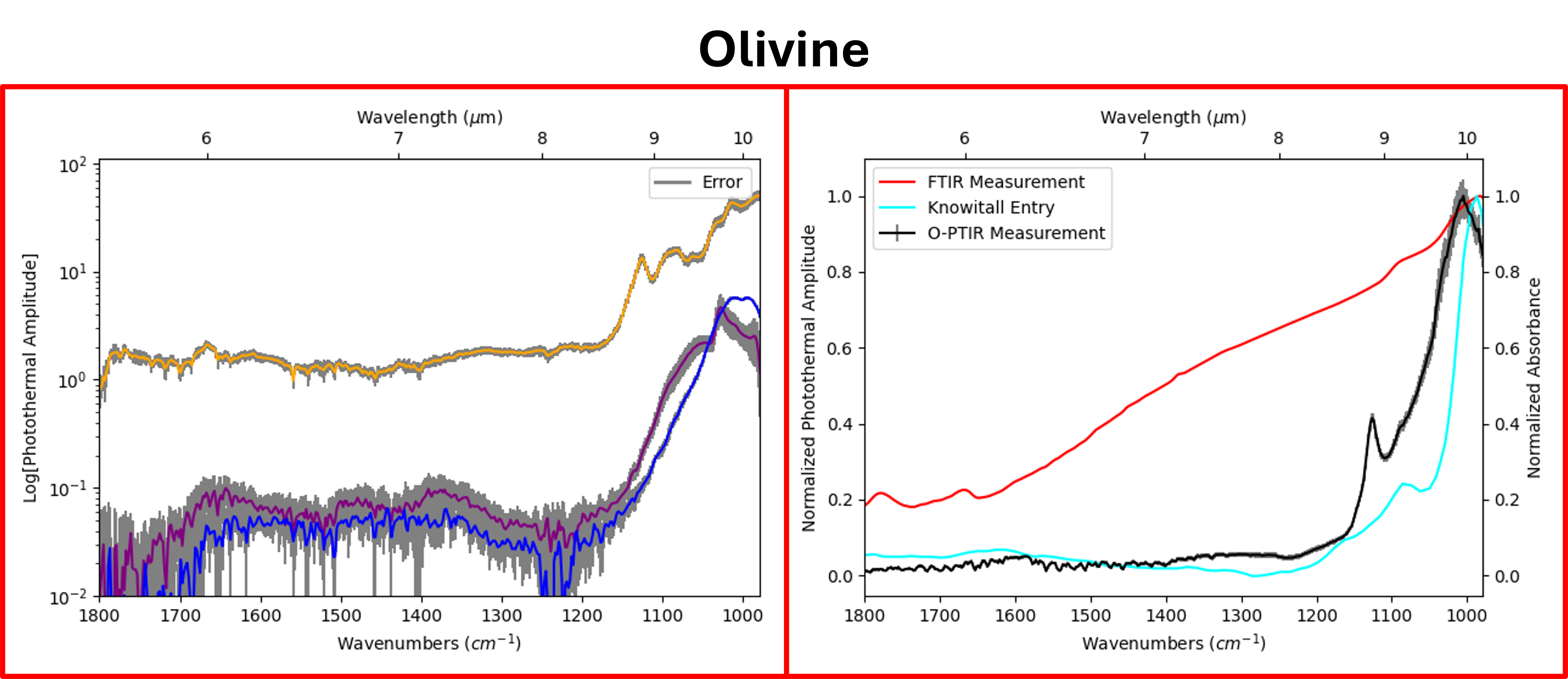}
    \caption{Mid-IR O-PTIR measurements of olivine (left) and the average spectrum from an O-PTIR hyperspectral map of olivine (right).}
    \label{f:olivine-figure}
\end{figure}

Figure \ref{f:olivine-figure} (left) shows O-PTIR measurements of olivine at chosen locations on the sample to show variations we hypothesize are due to grain orientation. Such a dependence on grain orientation has also been seen by \cite{taran2006octahedral} which specifically oriented grains of olivine and found differing spectral shapes and amplitudes. We similarly found that as different grains were measured, we saw differing spectral shapes and features as well as differing amplitudes in the spectra. However, the O-PTIR spectral measurements showed greater variance than those measurements seen in \cite{taran2006octahedral}.

Figure \ref{f:olivine-figure} (right) shows an O-PTIR measurement of olivine with standard error of mean error bars. Olivine shows O-PTIR peaks at 1006 and 1125 wavenumbers. The O-PTIR measurement is not identical to the FTIR measurement of the same sample nor the database entry in the Knowitall database. The only similarity of the FTIR measurement of the same sample and the O-PTIR measurement is the peak location. The similarity between the O-PTIR measurement and the Knowitall database entry is a better match. They share a similar spectral shape with peak locations slightly shifted. The less intense peak seems to be shifted more than the intense peak.

\cite{farmer1974ism} lists peaks associated with olivine in the IR. In the examined wavenumber range, olivine peaks at approximately 1000 $cm^{-1}$. This peak is only 6 wavenumbers displaced from one of the measured peaks in the O-PTIR.

\subsection{Smectite}

\begin{figure}[H]
    \centering
    \includegraphics[width=\textwidth]{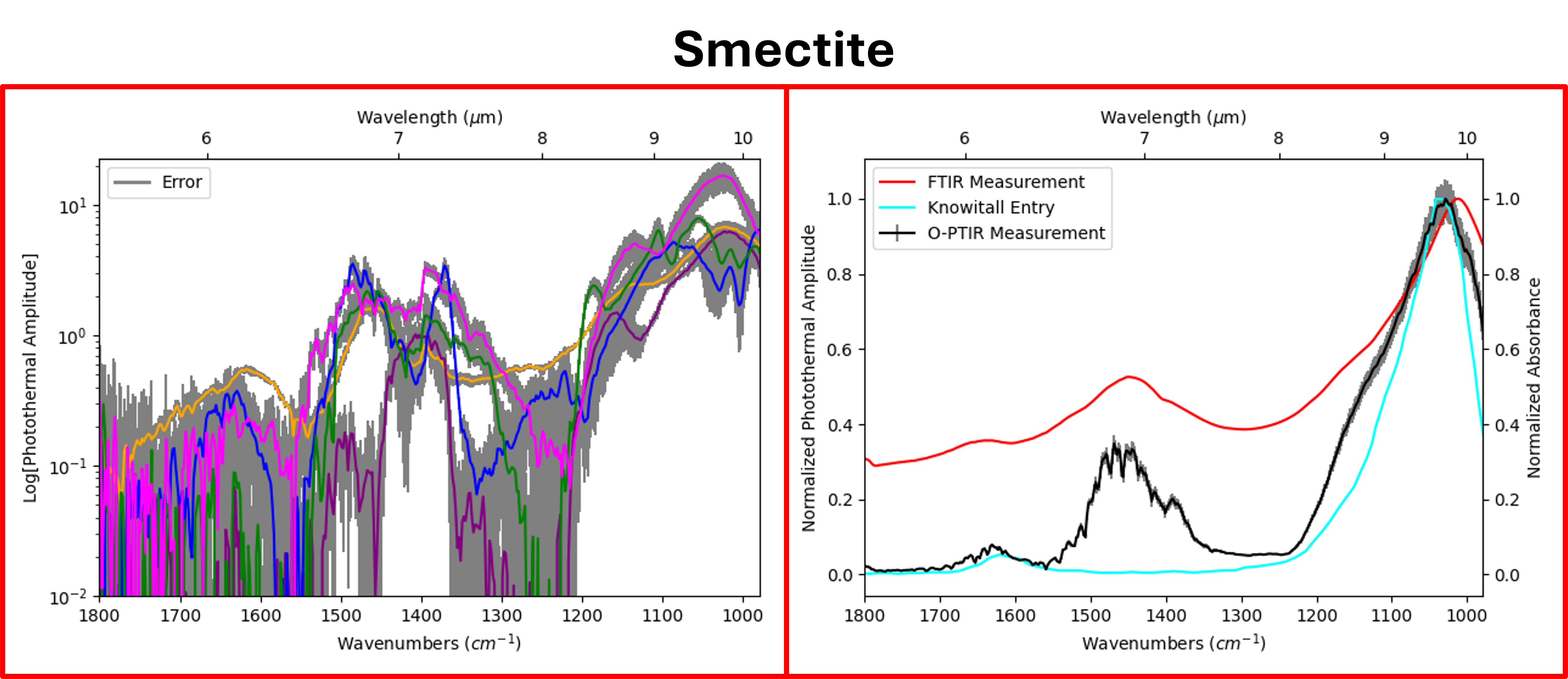}
    \caption{Mid-IR O-PTIR measurements of smectite (left) and the average spectrum from an O-PTIR hyperspectral map of smectite (right).}
    \label{f:smectite-figure}
\end{figure}

Figure \ref{f:smectite-figure} (left) shows O-PTIR measurements of smectite at chosen locations on the sample to show variations we hypothesize are due to grain orientation. 

Figure \ref{f:smectite-figure} (right) shows an O-PTIR measurement of smectite with standard error of mean error bars. Smectite shows O-PTIR peaks at 1028, 1447, and 1466 wavenumbers. The O-PTIR measurement has strong similarities with with both the FTIR measurement of the same sample nor the database entry in the Knowitall database. The FTIR measurement of the same sample shows a similar peak around 1050 $cm^{-1}$ but is slightly shifted to a lower wavenumber. However, the feature shows a similar width to the O-PTIR measurement. Additionally, the FTIR measurement also shares the peak between 1350 and 1550 $cm^{-1}$. The feature peaks very closely and had a nearly identical width. It also peaks between 1600 and 1700 $cm^{-1}$ though that feature is not as intense and is less certain than the other two features. The Knowitall database entry is missing the feature between 1350 and 1550 $cm^{-1}$, but does share the feature between 1600 and 1700 $cm^{-1}$ with a similar intensity as the O-PTIR measurement. The Knowitall database entry also shares a peak location with the O-PTIR measurement around 1050 wavenumbers, though the width of the feature is smaller there.

\section{Discussion}
\label{s:discuss}

\cite{cox2024photothermal} made a case for the use of the O-PTIR technique in planetary science space missions. We furthered the work by cataloguing and analyzing specific materials relevant to planetary science. In this section we will discuss the results of the measurements reported above. Primarily we will focus on the outcomes of individual point measurements and the averaged spectra of the hyperspectral maps.

\subsection{Granular Orientation Effects}
\label{s:granular-orientation-discussion}

Section \ref{s:granular-orientation} mentions the concern of granular orientation effects within our measurements. It also describes the measurements we took to attempt to demonstrate those effects in our samples. Section \ref{s:results} shows the granular orientation effects for each individual material as we were able to locate them. These effects are well known and studied in literature \citep[e.g.][]{serratosa1958determination,shuai2017quantitative}. These studies typically utilize single grains and change the orientation of the grain between each measurement. Sometimes these measurements have little to no effect. Other times these orientation changes cause reduction in peaks or even shifting peaks and changing spectral shapes.

In our measurements, we see that various grains in a bulk sample are capable of producing differing spectra. This is likely due to the current configuration of our system. The visible probing laser is circularly polarized, but the IR laser is linearly polarized. Typically, one of these possible spectra dominate the average over a large area. This could be due to random chance of which grains were in the measurement area. However, in most cases (e.g. bronzite, gypsum, hematite, etc.) certain spectral shapes and features peak at higher intensities. Those features are typically dominating the average spectrum. In some cases of multiple spectra, differing in orientation (e.g. anorthosite, magnesite, epsomite, etc.), it appears that the average spectrum is influenced by each of the found spectra.

A future work will take individual grains of materials and will analyze specifically how orientation effects spectra. This will better serve as a method to understand how granular orientation effects specifically affect O-PTIR measurements and clarify the role of light polarization in the instrument.

\subsection{Planetary Material Identification}
\label{s:material-identification}

In Section \ref{s:results} we report averaged spectra of hyperspectral maps for each material discussed in this work. These spectra all contain standard error of mean error bars. In all cases, there were no error bars that would significantly change the shape of a spectrum. This is significant as it demonstrates the consistency of the technique as it scans a sample. 

Further, the spectral shapes, spectral features, and spectral peaks are typically unique. In some cases the individual spectral measurements may have resembled other materials. However, in the case of bulk sample averages, the spectra appear unique. In other works \citep[such as][]{olson2020simultaneous,krafft2022optical}, Raman is used simultaneously with the O-PTIR technique. The simultaneous usage allows for even more unique features for each material. This is feasible due to the green visible laser causing fluorescence. Though the addition of Raman is out of the scope of this work, it does speak to the increased ability to identify planetary materials, which will be studied in future work.

\subsection{O-PTIR Comparison to IR Absorption}
\label{s:optir-compare}

The comparisons presented in Section \ref{s:results} showed that O-PTIR measurements consistently were comparable to IR absorption measurements. This was done by comparing O-PTIR measurements to FTIR measurements of the same sample and Wiley's Knowitall database entries of the same material. In some cases, these comparisons were nearly identical to the O-PTIR measurements. This is significant in that it shows that O-PTIR measurements are comparable to IR absorption measurements for minerals and granular materials. This shows that O-PTIR measurements produced in future planetary science missions will be comparable to existing IR absorption databases. Other works \citep[e.g.][]{klementieva2020super,bazin2022using,zhang2016depth} have made cases demonstrating the ability of O-PTIR to compare to IR absorption (in some cases even producing identical spectra). This work furthers this concept and demonstrates the ability to do the same for granular samples.

\section{Conclusion and Summary}
\label{s:conclusion}

In this work, we presented O-PTIR measurements of various materials relevant to planetary science. We furthered the work presented in \cite{cox2024photothermal} and expanded the presentation of the capabilities of the O-PTIR technique. We did so by: presenting the granular orientation effects on measurements through single point measurements, presenting analysis of bulk measurements of materials, and made direct comparisons of O-PTIR measurements to IR absorption measurements. We utilized the benefits of the O-PTIR technique described by \cite{cox2024photothermal} to produce a large catalogue of data for materials relevant to planetary science. This further demonstrates the ability of O-PTIR to benefit the planetary science community and the future of planetary science space missions.

In upcoming work, we will utilize the ability to identify minerals to identify constituent materials within a mixture. The ability to do this will be essential to understanding the origins of materials and to understanding the evolution  of these materials over time. These mixtures will be known regolith simulant mixtures and will be opaque materials. The ability to quantitatively analyze these materials will represent a step forward in analysis of foreign planetary materials.

\section*{Acknowledgments}

This work was supported by the NASA PICASSO grant \#80NSSC22K1231. Additionally, we would like to thank Dr. Kerri Donaldson Hanna for her fruitful conversations.

\bibliography{literature}{}
\bibliographystyle{elsarticle-harv}

\appendix

\section{FTIR Pellet Making Method}
\label{a:pellets}

Using PIKE Technologies' recipe\footnote{https://www.piketech.com/files/pdfs/PerfectPelletMakingAN.pdf} as a guide, a 200 mg pellet was made. The pellets were made with 1-2\% (\~2-4 mg) material of interest and 98-99\% (\~196-198 mg) KBr powder. The materials are ground into a fine powder and then mixed and transferred into a pellet press. The top anvil of the press is rotated clockwise with downward pressure to better level the sample before pressing the pellet. The press was evacuated of air for 1 minute and 30 seconds. 3 tons of force was applied to the pellet then immediately released. Then, 7 tons of force was applied for 30 seconds then immediately released. The press was evacuated again for 1 minute. Finally, 10 tons of force was applied for 2 minutes. The top of the press is disassembled from the base, the base is disconnected from the evacuation line, the rubber o-rings are removed from the press, and then the top of the press is placed back in the compactor. The compactor is used to push the pellet free of the press. The pellet is removed from the press and then placed into a holder for analysis. Finished pellets are placed in a dry box for storage until ready for analysis.

\section{Max Peaks and Deviations}

Below is a series of tables containing max peaks and deviations for the individually measured spectra for each material.

\begin{table}[H]
  \centering
  \caption{A table featuring the max peaks and max deviations of each O-PTIR measurement of anorthosite featured in Figure \ref{f:anorthosite-figure} (left).}
    \begin{tabular}{|l|r|r|}
    \toprule
    Spectrum & \multicolumn{1}{l|}{Max Peak} & \multicolumn{1}{l|}{Max Deviation} \\
    \midrule
    Spectrum 1 & 7.322693 & 0.324179 \\
    \midrule
    Spectrum 2 & 7.977738 & 0.213787 \\
    \midrule
    Spectrum 3 & 7.357513 & 0.314597 \\
    \bottomrule
    \end{tabular}%
  \label{t:anorthosite-table}%
\end{table}%

\begin{table}[H]
  \centering
  \caption{A table featuring the max peaks and max deviations of each O-PTIR measurement of basalt featured in Figure \ref{f:basalt-figure} (left).}
    \begin{tabular}{|l|r|r|}
    \toprule
    Spectrum & \multicolumn{1}{l|}{Max Peak} & \multicolumn{1}{l|}{Max Deviation} \\
    \midrule
    Spectrum 1 & 5.455014 & 0.195806 \\
    \midrule
    Spectrum 2 & 59.42456 & 3.613043 \\
    \bottomrule
    \end{tabular}%
  \label{t:basalt-table}%
\end{table}%

\begin{table}[H]
  \centering
  \caption{A table featuring the max peaks and max deviations of each O-PTIR measurement of bronzite featured in Figure \ref{f:bronzite-figure} (left).}
    \begin{tabular}{|l|r|r|}
    \toprule
    Spectrum & \multicolumn{1}{l|}{Max Peak} & \multicolumn{1}{l|}{Max Deviation} \\
    \midrule
    Spectrum 1 & 10.58403 & 0.693008 \\
    \midrule
    Spectrum 2 & 29.0392 & 1.997469 \\
    \midrule
    Spectrum 3 & 84.8512 & 5.504187 \\
    \midrule
    Spectrum 4 & 14.35081 & 0.534176 \\
    \bottomrule
    \end{tabular}%
  \label{t:bronzite-table}%
\end{table}%

\begin{table}[H]
  \centering
  \caption{A table featuring the max peaks and max deviations of each O-PTIR measurement of siderite featured in Figure \ref{f:siderite-figure} (left).}
    \begin{tabular}{|l|r|r|}
    \toprule
    Spectrum & \multicolumn{1}{l|}{Max Peak} & \multicolumn{1}{l|}{Max Deviation} \\
    \midrule
    Spectrum 1 & 2.905344 & 0.137475 \\
    \midrule
    Spectrum 2 & 2.974705 & 0.157269 \\
    \midrule
    Spectrum 3 & 4.072176 & 0.541465 \\
    \bottomrule
    \end{tabular}%
  \label{t:siderite-table}%
\end{table}%

\begin{table}[H]
  \centering
  \caption{A table featuring the max peaks and max deviations of each O-PTIR measurement of ferrihydrite featured in Figure \ref{f:ferrihydrite-figure} (left).}
    \begin{tabular}{|l|r|r|}
    \toprule
    Spectrum & \multicolumn{1}{l|}{Max Peak} & \multicolumn{1}{l|}{Max Deviation} \\
    \midrule
    Spectrum 1 & 1.227325 & 0.234679 \\
    \bottomrule
    \end{tabular}%
  \label{t:ferrihydrite-table}%
\end{table}%

\begin{table}[H]
  \centering
  \caption{A table featuring the max peaks and max deviations of each O-PTIR measurement of gypsum featured in Figure \ref{f:gypsum-figure} (left).}
    \begin{tabular}{|l|r|r|}
    \toprule
    Spectrum & \multicolumn{1}{l|}{Max Peak} & \multicolumn{1}{l|}{Max Deviation} \\
    \midrule
    Spectrum 1 & 2.681214 & 0.101136 \\
    \midrule
    Spectrum 2 & 2.453152 & 0.142794 \\
    \midrule
    Spectrum 3 & 1.9243 & 0.071491 \\
    \bottomrule
    \end{tabular}%
  \label{t:gypsum-table}%
\end{table}%

\begin{table}[H]
  \centering
  \caption{A table featuring the max peaks and max deviations of each O-PTIR measurement of hematite featured in Figure \ref{f:hematite-figure} (left).}
    \begin{tabular}{|l|r|r|}
    \toprule
    Spectrum & \multicolumn{1}{l|}{Max Peak} & \multicolumn{1}{l|}{Max Deviation} \\
    \midrule
    Spectrum 1 & 9.201209 & 0.458004 \\
    \midrule
    Spectrum 2 & 0.609176 & 0.091371 \\
    \midrule
    Spectrum 3 & 3.38826 & 0.114438 \\
    \bottomrule
    \end{tabular}%
  \label{t:hematite-table}%
\end{table}%

\begin{table}[H]
  \centering
  \caption{A table featuring the max peaks and max deviations of each O-PTIR measurement of hydrated silica featured in Figure \ref{f:hydratedsilica-figure} (left).}
    \begin{tabular}{|l|r|r|}
    \toprule
    Spectrum & \multicolumn{1}{l|}{Max Peak} & \multicolumn{1}{l|}{Max Deviation} \\
    \midrule
    Spectrum 1 & 20.57146 & 0.598685 \\
    \midrule
    Spectrum 2 & 19.84694 & 3.233245 \\
    \midrule
    Spectrum 3 & 17.09244 & 2.493837 \\
    \bottomrule
    \end{tabular}%
  \label{t:hydratedsilica-table}%
\end{table}%

\begin{table}[H]
  \centering
  \caption{A table featuring the max peaks and max deviations of each O-PTIR measurement of ilmenite featured in Figure \ref{f:ilmenite-figure} (left).}
    \begin{tabular}{|l|r|r|}
    \toprule
    Spectrum & \multicolumn{1}{l|}{Max Peak} & \multicolumn{1}{l|}{Max Deviation} \\
    \midrule
    Spectrum 1 & 0.706231 & 0.114065 \\
    \midrule
    Spectrum 2 & 0.861958 & 0.188545 \\
    \midrule
    Spectrum 3 & 4.23806 & 0.157548 \\
    \bottomrule
    \end{tabular}%
  \label{t:ilmenite-table}%
\end{table}%

\begin{table}[H]
  \centering
  \caption{A table featuring the max peaks and max deviations of each O-PTIR measurement of magnetite featured in Figure \ref{f:magnetite-figure} (left).}
    \begin{tabular}{|l|r|r|}
    \toprule
    Spectrum & \multicolumn{1}{l|}{Max Peak} & \multicolumn{1}{l|}{Max Deviation} \\
    \midrule
    Spectrum 1 & 12.28606 & 0.353978 \\
    \bottomrule
    \end{tabular}%
  \label{t:magnetite-table}%
\end{table}%

\begin{table}[H]
  \centering
  \caption{A table featuring the max peaks and max deviations of each O-PTIR measurement of magnesite featured in Figure \ref{f:magnesite-figure} (left).}
    \begin{tabular}{|l|r|r|}
    \toprule
    Spectrum & \multicolumn{1}{l|}{Max Peak} & \multicolumn{1}{l|}{Max Deviation} \\
    \midrule
    Spectrum 1 & 22.60104 & 2.37159 \\
    \midrule
    Spectrum 2 & 26.57293 & 0.742645 \\
    \midrule
    Spectrum 3 & 9.402586 & 0.405154 \\
    \midrule
    Spectrum 4 & 5.560586 & 0.844084 \\
    \bottomrule
    \end{tabular}%
  \label{t:magnesite-table}%
\end{table}%

\begin{table}[H]
  \centering
  \caption{A table featuring the max peaks and max deviations of each O-PTIR measurement of epsomite featured in Figure \ref{f:epsomite-figure} (left).}
    \begin{tabular}{|l|r|r|}
    \toprule
    Spectrum & \multicolumn{1}{l|}{Max Peak} & \multicolumn{1}{l|}{Max Deviation} \\
    \midrule
    Spectrum 1 & 1.113239 & 0.282312 \\
    \midrule
    Spectrum 2 & 2.566355 & 0.399092 \\
    \midrule
    Spectrum 3 & 6.026462 & 0.246884 \\
    \bottomrule
    \end{tabular}%
  \label{t:epsomite-table}%
\end{table}%

\begin{table}[H]
  \centering
  \caption{A table featuring the max peaks and max deviations of each O-PTIR measurement of olivine featured in Figure \ref{f:olivine-figure} (left).}
    \begin{tabular}{|l|r|r|}
    \toprule
    Spectrum & \multicolumn{1}{l|}{Max Peak} & \multicolumn{1}{l|}{Max Deviation} \\
    \midrule
    Spectrum 1 & 95.26071 & 9.271133 \\
    \midrule
    Spectrum 2 & 4.663567 & 1.491738 \\
    \midrule
    Spectrum 3 & 5.732936 & 0.330583 \\
    \bottomrule
    \end{tabular}%
  \label{t:olivine-table}%
\end{table}%

\begin{table}[H]
  \centering
  \caption{A table featuring the max peaks and max deviations of each O-PTIR measurement of smectite featured in Figure \ref{f:smectite-figure} (left).}
    \begin{tabular}{|l|r|r|}
    \toprule
    Spectrum & \multicolumn{1}{l|}{Max Peak} & \multicolumn{1}{l|}{Max Deviation} \\
    \midrule
    Spectrum 1 & 6.752547 & 0.922227 \\
    \midrule
    Spectrum 2 & 6.23368 & 1.048889 \\
    \midrule
    Spectrum 3 & 6.443592 & 1.591813 \\
    \midrule
    Spectrum 4 & 7.860179 & 2.305671 \\
    \midrule
    Spectrum 5 & 16.75025 & 4.600344 \\
    \bottomrule
    \end{tabular}%
  \label{t:smectite-table}%
\end{table}%

\end{document}